\documentclass[%
 aip,
 amsmath,amssymb,
 reprint,
]{revtex4-1}

\usepackage{graphicx}% Include figure files
\usepackage{dcolumn}% Align table columns on decimal point
\usepackage{bm}% bold math
\usepackage[utf8]{inputenc}
\usepackage[T1]{fontenc}
\usepackage{mathptmx}
\usepackage{xfrac}
\usepackage[table,xcdraw]{xcolor}
\usepackage[normalem]{ulem}
\useunder{\uline}{\ul}{}
\usepackage{enumitem}

\begin{document}

\preprint{AIP/123-QED}

\title[A scalable helium gas cooling system for trapped-ion applications]{A scalable helium gas cooling system for trapped-ion applications}
% Force line breaks with \\

\author{F. R. Lebrun-Gallagher}
\affiliation{Sussex Centre for Quantum Technologies, University of Sussex, Brighton, BN1 9RH, U.K.}
\author{N. Johnson}%
\affiliation{Sussex Centre for Quantum Technologies, University of Sussex, Brighton, BN1 9RH, U.K.}
\author{M. Akhtar}
\affiliation{Sussex Centre for Quantum Technologies, University of Sussex, Brighton, BN1 9RH, U.K.}
\affiliation{Universal Quantum Ltd, Brighton, BN1 6SB, U.K.}
\author{S. Weidt}
\affiliation{Sussex Centre for Quantum Technologies, University of Sussex, Brighton, BN1 9RH, U.K.}
\affiliation{Universal Quantum Ltd, Brighton, BN1 6SB, U.K.}
\author{D. Bretaud}
\affiliation{Sussex Centre for Quantum Technologies, University of Sussex, Brighton, BN1 9RH, U.K.}
\affiliation{QOLS, Blackett Laboratory, Imperial College London, London, SW7 2BW, U.K.}
\author{S. J. Hile}
\affiliation{Sussex Centre for Quantum Technologies, University of Sussex, Brighton, BN1 9RH, U.K.}
\author{A. Owens}
\affiliation{Sussex Centre for Quantum Technologies, University of Sussex, Brighton, BN1 9RH, U.K.}
\author{W. K. Hensinger}
\email{w.k.hensinger@sussex.ac.uk}
\affiliation{Sussex Centre for Quantum Technologies, University of Sussex, Brighton, BN1 9RH, U.K.}
\affiliation{Universal Quantum Ltd, Brighton, BN1 6SB, U.K.}

\date{\today}

\begin{abstract}

Microfabricated ion-trap devices offer a promising pathway towards scalable quantum computing. Research efforts have begun to focus on the engineering challenges associated with developing large-scale ion-trap arrays and networks. However, increasing the size of the array and integrating on-chip electronics can drastically increase the power dissipation within the ion-trap chips. This leads to an increase in the operating temperature of the ion-trap and limits the device performance. Therefore, effective thermal management is an essential consideration for any large-scale architecture. Presented here is the development of a modular cooling system designed for use with multiple ion-trapping experiments simultaneously. The system includes an extensible cryostat that permits scaling of the cooling power to meet the demands of a large network. Following experimental testing on two independent ion-trap experiments, the cooling system is expected to deliver a net cooling power of 111~W at $\sim$70~K to up to four experiments. The cooling system is a step towards meeting the practical challenges of operating large-scale quantum computers with many qubits.

\end{abstract}

\maketitle

\section{\label{sec: Introduction} Introduction}

Trapped ions are a leading platform for quantum computers (QCs) \cite{Bruzewicz2019}, quantum simulators \cite{Blatt2012}, high-precision quantum sensors \cite{Biercuk2010, Goldstein2011} and fundamental physics research \cite{Massar2015, Horvath1997}. A number of trapped-ion QC architectures have been proposed which focus on developing scalable designs for future fault-tolerant devices with many qubits \cite{Lekitsch2017, Monroe2014}. Microfabricated ion-traps\cite{Romaszko2020} are a promising architecture for a modular QC\cite{Lekitsch2017}. A recently published QC blueprint \cite{Lekitsch2017} uses modular surface ion-traps, with integrated on-chip electronics for ion transport, quantum state manipulation and readout. Each addition of these on-chip features, however, increases the power dissipated by components such as radio frequency (RF) conductors, digital-to-analogue converters (DACs), fluorescence detectors, integrated ovens, and magnetic field generating structures. The thermal management of the modular QC is then essential to retain a reasonable operating temperature and also to improve ion-trap performance \cite{Labaziewicz2008}. Operation at cryogenic temperatures suppresses electric field noise \cite{Labaziewicz2008}, leading to a reduction in ion motional heating \cite{Bruzewicz2015} and significantly improves trapping lifetimes, qubit coherence times and quantum logic gate fidelities \cite{Wineland1998}. 

When considering a target operating temperature for an ion-trap, one of the primary concerns is the motional heating rate of the ion. Motional heating is driven by electric field fluctuations within the surface layers of a chip \cite{Wineland1998}. The ion motional heating rate $\dot{n}$ follows a temperature dependency of the form \cite{Wineland1998} 

\begin{equation}
\dot{n} \propto 1+ \left(\frac{T}{T_0} \right)^{\beta},
\end{equation}

\noindent where $T_0$ is the thermal activation temperature, found to be in the 10--75~K temperature range \cite{Labaziewicz2008, Wineland1998}, $T$ is the temperature of the ion-trap and $\beta$ is an exponent describing the high temperature dependency, typically in the 1.5--4.1 range\cite{Labaziewicz2008, Bruzewicz2015}. This indicates that cooling below the thermal activation temperature affords only a marginal reduction in the ion motional heating rate. In practice most of the gains are made when cooling from 300~K to 100~K, leading to a heating rate reduction of several orders of magnitude \cite{Labaziewicz2008, Wineland1998, Deslauriers2006}. 

Operation at temperatures below 100~K still may be desirable, since it decreases the vacuum pressure and therefore reduces the probability of background gas molecules colliding with the ion. For example, systems operating at 4~K have lower vacuum pressures by about 4 orders of magnitude when compared to the typical vacuum pressure of room-temperature experiments \cite{Brandl2016}. However, operation at 4~K presents several technical challenges, such as with implementing silicon-based electronics \cite{Ghibaudo1997, Rahman2016}.

This paper describes the development of a scalable cooling system with novel application to ion-trap devices and other high power experiments. Designed as a step towards a scalable cryogenic cooling solution suited to the QC architecture described by Lekitsch et al. \cite{Lekitsch2017}, the system uses circulating helium gas to cool multiple ion-trapping experiments independently to below 70~K.

\section{\label{sec: Experimental Set-up} Design Overview}

\begin{figure*}
    \centering
    \includegraphics{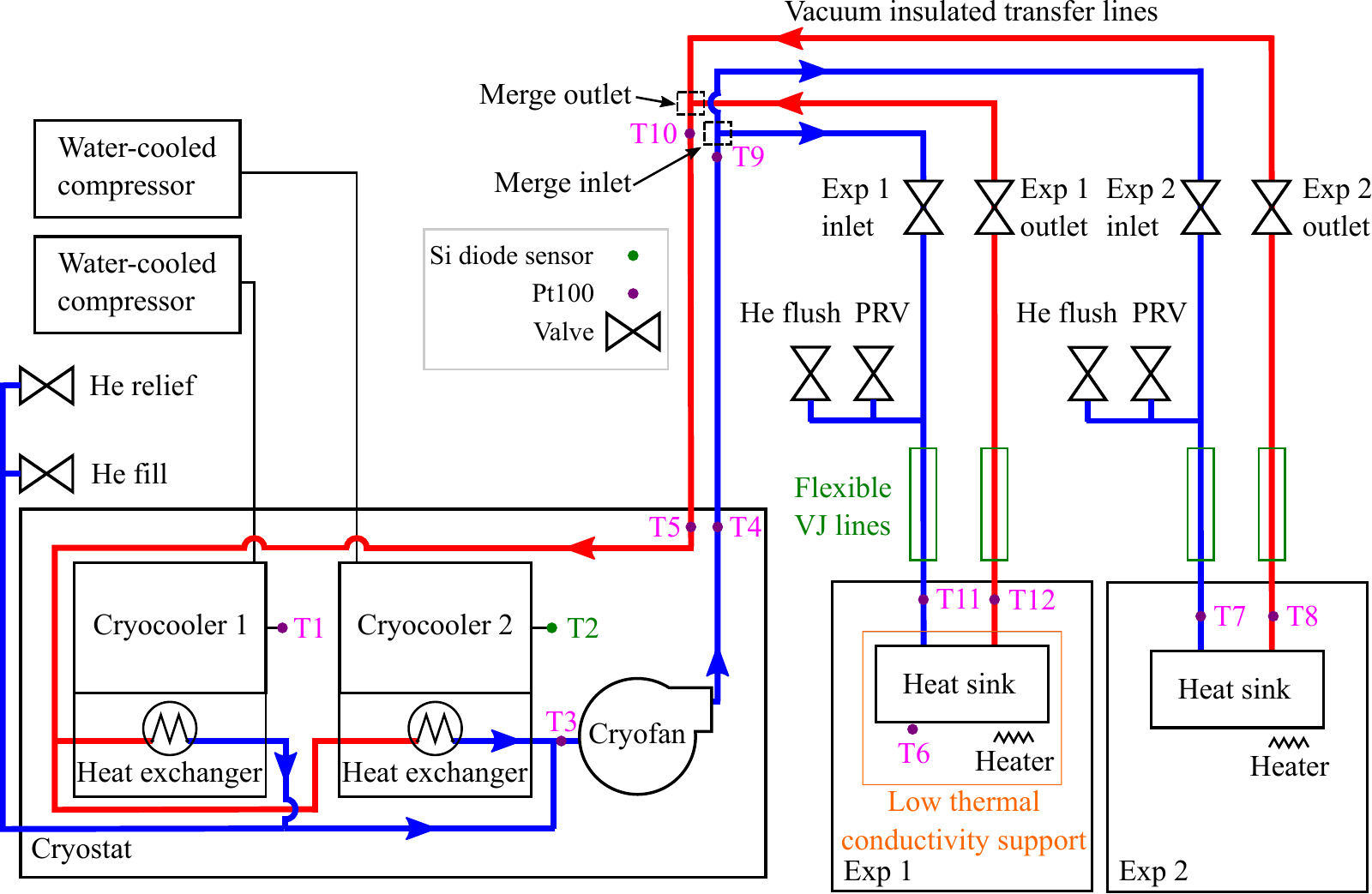}
    \caption{Schematic overview of the cryogenic cooling system. Cryocooler 1 (Gifford-McMahon CH110) and cryocooler 2 (Gifford-McMahon AL330) are remotely connected to two ion-trap experiments: `Exp 1' and `Exp 2'. Not shown are the lines that can be used to connect two further experiments in parallel at the merge inlet and merge outlet points (in a similar fashion to Exp 2.). At these points, the division of the helium gas flow is controlled by the manual cryogenic valves allocated to each experiments. The in-vacuum support structure that is used for Exp 1 is detailed within the text. The Pt100 sensor T1 and the silicon diode sensor T2 are fitted to the cold finger of cryocooler 1 and 2 respectively and are selected to best match the thermal performance of each cryocooler.}
    \label{fig: Cryo System}
\end{figure*}

A range of cryogenic cooling solutions have been demonstrated with small-scale ion-trapping experiments. `Wet' systems consist of either an open bath \cite{Poitzsch1996, Deslauriers2006, Antohi2009} or an open continuous flow cryostat \cite{Brandl2016} that use 4~K liquid helium or 77~K liquid nitrogen as coolant. Cryogenic temperatures are sustained for the duration of the liquid cryogen reserve which is lost to the environment by evaporation. `Dry' cryogenic systems do not require replenishment of cryogen, which is an attractive feature. Instead, cooling is provided by a closed cryogen-free refrigeration cycle. These have found wide application to ion trapping experiments in the form of Pulse-Tube \cite{Schwarz2012, Vittorini2013} or Gifford-McMahon (GM) \cite{Sage2012, Pagano2018} cryocoolers. However, size constraints prevent sufficient numbers of individual, commercially available cryostats from being used together within closely-spaced ion-trap arrays. Therefore, a new cooling architecture that avoids these constraints is proposed. The architecture is based on a closed-loop cryogenic helium gas circulation system; separating the cryocooler from the ion-traps in a way which is in principle extensible to accommodate large numbers of both ion-trap elements and cryocooler cold-heads.

A cooling system based on this architecture was constructed to interface between four independent ion-trap experiments. The cooling system consists of three sections: a main cryostat chamber that houses two single-stage Gifford-McMahon cryocooler cold heads and a centrifugal fan; a network of vacuum-jacketed (VJ) helium transfer lines that distribute the cold gas from the cryostat to the ion-trap experiments, and a cryogenic heat sink installed on each experiment. A diagram of the cooling system is shown in Figure \ref{fig: Cryo System}. The cooling system distributes helium gas to multiple experiments configured in parallel. Individual experiments can be connected and disconnected from the cooling circuit independently. Together, the two cryocoolers deliver a total cooling power of 396~W at 80~K \cite{AL330,CH110}.

Inside the cryostat chamber, each cold head interfaces with a heat exchanger, which is braised to the stainless steel feedthroughs carrying helium gas. This minimises the thermal resistance between the cold head and the helium cryogen. The compressors used to operate the cryocoolers are located in a separate room next to the laboratory to minimise vibration at the ion-trap experiments.

Helium gas is circulated using a Stirling Cryogenics Noordenwind centrifugal fan (or ‘cryofan’). The flow of helium gas is routed to the two cold heads in parallel and then recombined prior to the inlet of the cryofan. This parallel circuit configuration reduces the overall flow resistance across the cryostat section, while allowing for faster thermalisation of the gas through the heat exchangers. In this configuration the total available cooling power can be scaled by integrating extra cryocoolers in parallel. Within the cryostat vessel, leak-tight connections are made using both vacuum coupling radiation (VCR) fittings and by direct welding. To assist in dampening vibrations caused by the cryocooler, a section of flexible, mesh-reinforced, stainless steel braided hose was installed at the cryofan inlet.

The cryostat vessel and the VJ lines are maintained under vacuum (< 10$^{-4}$~mbar) to prevent conductive heat leaks. The cold heads, cryofan and stainless-steel tubing are lined with 30 layers of multilayer insulation (MLI) to minimise radiative heat leaks. Overall, the transfer lines are expected to contribute to thermal losses by $\sim$0.2~W/m across the network, which extends for $\sim$30~m. 

At the locations of each ion-trapping experiment, the VJ transfer lines separate into two individual supply and return lines. These are constructed from flexible VJ tubing to further attenuate vibrations from the cryostat and correct for thermal expansion mismatches between sections of different temperatures. Cryogenic vacuum barriers are fitted between the main rigid VJ line and the flexible sections, and on the helium feedthroughs of the ion-trap vacuum chamber. This allows for the flexible sections to be disconnected from the vacuum chamber, while maintaining the vacuum insulation within the main VJ transfer line and the UHV (ultra-high vacuum) within the ion-trap experiment.

The system is pressurised to 20~bar and the overall helium gas mass flow rate is controlled by adjusting the rotation speed of the cryofan impeller which can be throttled between 6,000 and 21,000 revolutions per minute (rpm). A schematic of the helium gas transfer network, which connects the cryostat to the ion-trap experiments, is shown in Figure \ref{fig: Cryo System}.

The local flow of helium gas allocated to each ion-trapping experiment is regulated using two manual cryogenic valves which are fitted on the feed and return transfer lines. When an experiment is disconnected, the cold pressurised helium gas trapped in the experiment heat sink warms up and expands. A cryogenic safety pressure release valve (PRV), set to 23~bar, is fitted at the experiment return VJ line to avoid over-pressure in the heat sink during thermalisation to room temperature. An additional valve (`He flush' valve in Figure \ref{fig: Cryo System}) connects the inner helium line to atmosphere and allows for each experiment to be evacuated and flushed with helium before being cooled down. Flushing with helium removes any trapped air and water vapour that can freeze on the impeller blades of the cryofan, or inside the heat sink, which may cause damage or choke the helium flow.

\section{\label{sec: Interfacing with an Ion -Trap Experiment} Interfacing with an Ion -Trap Experiment}

\begin{figure*}
    \centering
    \includegraphics[width=\textwidth]{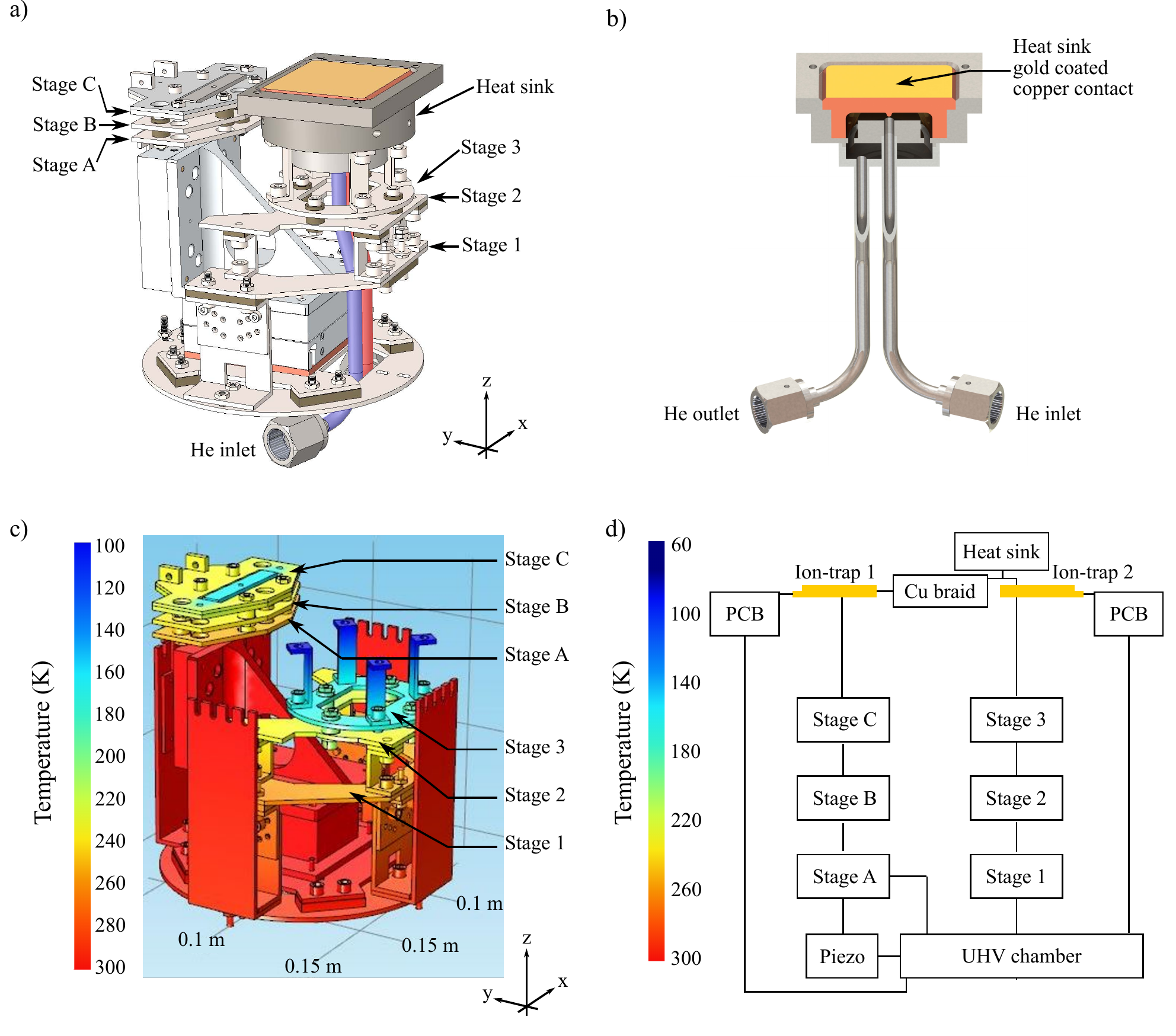}
    \caption{a) Diagram of the in-vacuum, low thermal conductivity support structure for two ion-traps (not shown) that are part of Exp 1. The three stages that support the first ion-trap (A--C), and similarly for a second ion-trap (1--3), are designed to provide levels of thermal resistance between the ion-traps and the rest of the support structure. For clarity, an overlay of blue has been added to the helium inlet tubing and red for the helium outlet tubing. b) Sectional view of the heat sink used to interface an ion-trap with the cooling system. c) Illustration of the thermal distribution when the cooling system is engaged. d) Schematic of the thermal paths between the different sections in the low thermal conductivity support structure and mounted ion-traps.}
    \label{fig: In-Vacuum Mounting}
\end{figure*}

In each ion-trap experiment, active heat loads arise from various sources, e.g. RF losses within the substrate of the microchips, Ohmic heating within microfabricated wires carrying electrical currents\cite{Lekitsch2017, Ospelkaus2011}, and power dissipation from chip-integrated electronics\cite{Stuart2019}. Passive heat loads result from both thermal conduction and radiation. Conductive losses occur along solid paths provided by feedthrough connections into the vacuum chamber and by the in-vacuum mounting structure. Radiative heat transfer occurs between the room temperature vacuum apparatus and the cryogenic surfaces that are in direct line of sight.

To minimise the passive thermal load in experiment 1 (Exp 1 in Figure \ref{fig: Cryo System}) a high stability, low thermal conductivity support structure was designed with a large thermal resistance. The design can be seen in Figure \ref{fig: In-Vacuum Mounting}a. The mounting structure is intended to support two ion-trap chips which are not shown in the figure. One ion-trap would be mounted to the heat sink and a second ion-trap would be mounted on the adjacent structure with cooling provided by a flexible copper braid that thermally anchors it to the first.

The heat sink that interfaces an ion-trap with the cooling system can be seen in Figure \ref{fig: In-Vacuum Mounting}b. It consists of a compact, stainless steel/OFHC copper enclosure within which cold helium gas is circulated. The heat sink uses a jet-impingement design where the pressurised helium jet is directed onto a copper plate, which is used as the heat transfer interface (see Figure \ref{fig: Cryo System}b). This approach generates only a small pressure drop over a large range of flow rates. Alternative heat sink designs, such as ones using pins, fins or microchannels may deliver improved heat transfer performance but carry the risk of a larger pressure drop penalty \cite{Kandlikar2005} while also being more complex to manufacture \cite{Dixit2015}. These designs remain under consideration for future upgrades of the system. In order to interface with the ion-trap, the heat sink copper plate is coated with 3~$\mu$m of gold and an additional copper support is added onto which the trap is then die-bonded. The gold-plated copper to copper structure ensures good thermal anchoring with an expected thermal contact resistance of 1.1~W/(K$\cdot$cm$^2$) \cite{Salerno1997, Mykhaylyk2012}. A 50~$\Omega$ cryogenic heater, capable of continuous operation under 1~A, is used to stabilise the heat sink temperature through a control loop and for warm-up.

The support structure around the heat sink reduces the conductive and radiative heat transfer from the surrounding vacuum chamber, effectively minimising the passive heat load on the heat sink. The structure is subdivided into three stages operating at different temperatures. Each stage of the assembly is built with grade 5 titanium alloy (Ti6Al4V) owing to its low thermal conductivity ranging from 1 to 3~W/(m$\cdot$K), and low coefficient of thermal expansion in the 20--100~K temperature range\cite{Marquardt2002}. These are respectively one third and one half of that of stainless steel over the same temperature range \cite{Marquardt2002}. An approach to further increase the thermal insulation of the support structure is to prevent metal-to-metal contacts between the cryogenic support stages. For this purpose, UHV compatible PEEK spacers were used between stages to minimise conductive losses as they increase the overall thermal path while having a low thermal conductivity at cryogenic temperatures \cite{Gottardi2001}. In addition, the thermal load from thermal radiation is also mitigated via the electromechanical polishing of the titanium parts which minimises emissivity. A three-dimensional, finite element method simulation of the conductive heat transfer through the support structure geometry was carried out using COMSOL and accounts for the thermal conductivity of the materials used. A visualisation of the temperature distribution on the support structure is provided in Figure \ref{fig: In-Vacuum Mounting}c. Overall, the structure exhibits a total thermal resistance of 146~K/W between the ion-trap and the vacuum chamber. A schematic of the thermal paths between the vacuum chamber and the ion-traps can be seen in Figure \ref{fig: In-Vacuum Mounting}d.

Thermometry is provided using platinum resistance temperature sensors (Pt100s). Each sensor is thermally anchored using a spring loaded copper clamp. The electrical connections to each sensor are made using long, phosphor-bronze cables which have reduced thermal conductivity compared to copper.

\section{\label{sec: Testing and Characterisation} Testing and Characterisation}

The performance of the cooling system was evaluated in a set of cryogenic tests that were carried out first with a single ion-trap experiment, and then with two experiments connected in parallel to the cooling system. The results in the following sections demonstrate the minimum operating temperature of the system, the time to reach a steady state, and the frequency and amplitude of vibrations at the ion-trap locations.

\subsection{Cooling of a single ion-trap experiment under no \\active heat load}\label{sec: Cooling 1 Exp No Heat Load}

\begin{figure}
    \centering
    \includegraphics{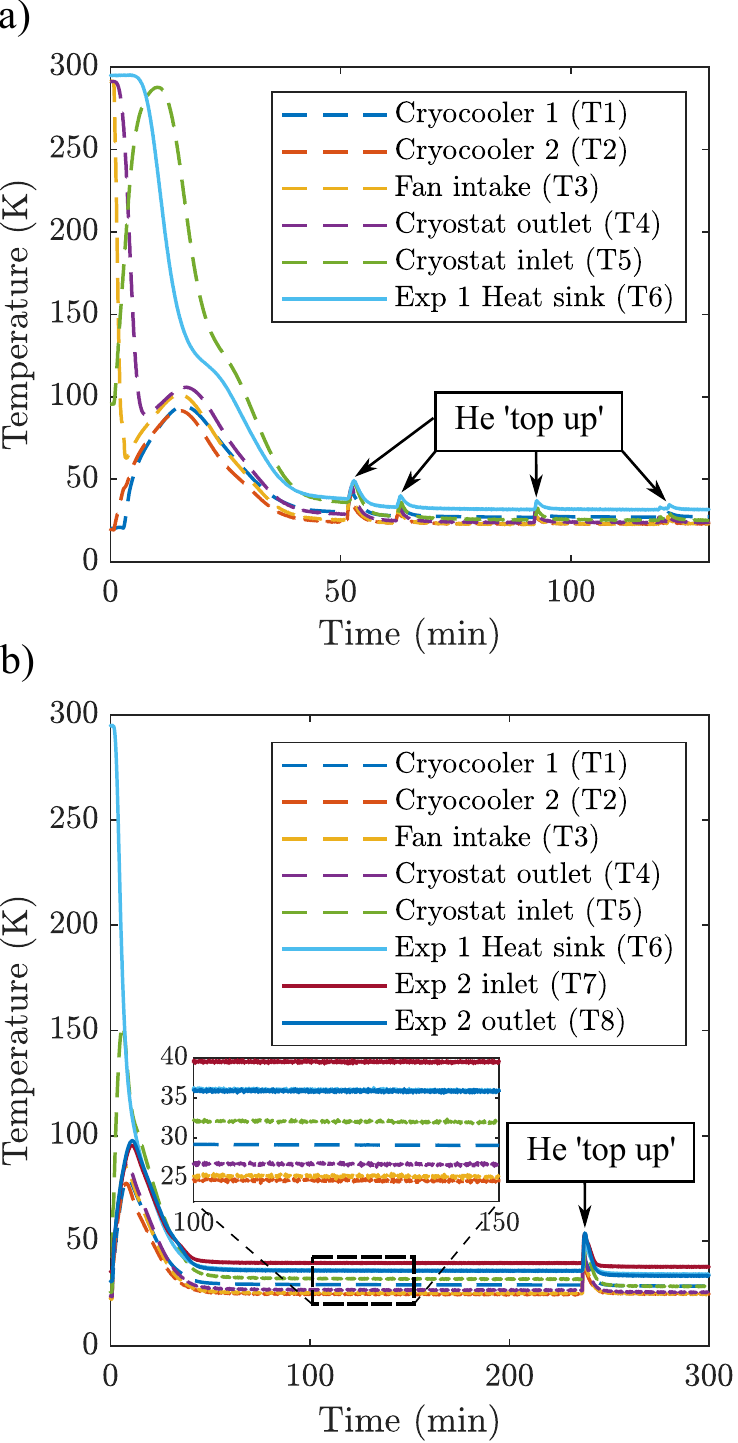}
    \caption{Temperature profiles of a) turning on the cryogenic cooling system with a single experiment (Exp 1) and b) during the connection of an additional experiment (Exp 1) in parallel to an already cooled experiment (Exp 2). The locations of the temperature sensors T1--T8 can be found on Figure \ref{fig: Cryo System}. The graph inset shows a subset of the data for clarity.}
    \label{fig: Temperature}
\end{figure}

The cooling system was connected to a single ion-trap experiment (Exp 1 in Figure \ref{fig: Cryo System}) after both cryocoolers were allowed to reach a base temperature of 20~K. The system was filled with helium gas and pressurised to 20~bar. To circulate helium to the experiment, the valves controlling the helium flow to the heat sink at the ion-trap experiment were opened. The cryofan, which was initially off, was switched on and set at its maximum rotation speed of 21,000 rpm. Figure \ref{fig: Temperature}a shows the evolution of the temperature of the cooling circuit when an experiment at room-temperature is connected. For this test, Pt100 temperature sensors were installed into the system as indicated by Figure \ref{fig: Cryo System}. Helium gas circulates from the cold head heat exchangers to the cryofan and exits the cryostat vessel at the maximum flow rate. This resulted in cryogenic temperatures ($\leq$100~K) measured at the cryofan intake and the cryostat outlet locations $\sim$5~min after cryofan activation.

As the cold helium gas circulates through the transfer lines to the heat sink at the ion-trap experiment, it displaces room temperature gas within the lines. As a consequence, the cryostat inlet sensor measures a transient rise in temperature. This effect was also measured to a lesser extent by the other sensors throughout the circuit. After $\sim$15~min, the rising temperature reaches a maximum and begins to decrease. Within 30~min of activating the cryofan, all sensor readings were below 40~K.

Since the helium pressure decreases with temperature, additional `topping-up' of gas was required to maintain a 20~bar operating pressure. These top-ups were performed four times over the course of $\sim$90~min, until the temperature and pressure stabilised. The maximum cooling rate at the experiment heat sink was measured to be 23~K/min. Full control of the cooling rate is also possible by manually adjusting the inlet and return helium valves. The highest cooling rate of all sensor locations was 150~K/min, measured at the cryofan intake. A steady-state base temperature of 32~K was reached at the experiment heat sink after 3~hours of operation under no active heat load.

\subsection{Cooling of two ion-trap experiments under no active heat load}\label{sec: Cooling 2 Exp No Heat Load}

This section investigates how connecting and disconnecting an experiment to the cooling system influences its overall performance. For this purpose, Exp 1 was connected into the cooling system circuit following cooling of Exp 2 to 32~K. The temperature evolution of Exp 1 is shown in Figure \ref{fig: Temperature}b.

The cryogenic helium gas valves were opened at Exp 1 to allow cold helium gas at 32~K to circulate. For Exp 2, in place of a heat sink, a length of \sfrac{1}{4}$\,$" copper pipe was installed with two Pt100 sensors at the inlet and outlet of the copper fitting. As the valves to Exp 1 were opened a transient temperature rise was observed. At Exp 2 the temperature rises for 11~min until a maximum temperature of 97~K is reached, before decreasing again to a base temperature of 36~K 60~min after the start of the test sequence. At Exp 1, the heat sink also reaches 36~K after 60~min. A single helium gas top-up to 20~bar was performed. This led to a temporary spike in temperature, before settling to a steady state temperature of 34~K at both experiments.

Under no active load, the steady-state temperature at the outlet is expected to be marginally higher than that of the inlet owing to the passive thermal leaks to the experiment heat sink. However, the inlet temperature is measured to be higher than the outlet. This effect is attributed to unequal thermal anchoring of the sensors to the helium lines and the passive heat load from connecting wires.

\subsection{Cooling power}\label{sec: Cooling Power}

\begin{figure}
    \centering
    \includegraphics{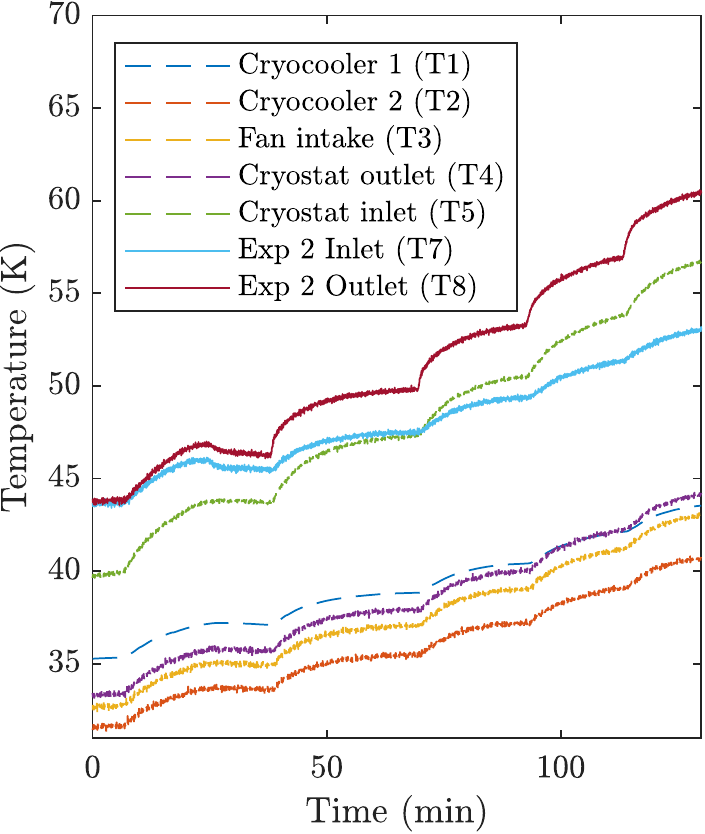}
    \caption{Temperature profiles of Exp 2 for increasing heater power in steps of 10~W from 40~W to 80~W. Note that (as in Figure \ref{fig: Temperature}) there is poor thermal anchoring of T7.}
    \label{fig: Cooling Power}
\end{figure}

This section investigates the relative contribution of active and passive heat loads in the cooling circuit. At thermal steady-state, the total cooling power of the cryogenic system is equal to the sum of the active and passive heat loads.

The cooling power was measured directly by applying an electric current to a heater installed within each ion-trap experiment (see Figure \ref{fig: Cryo System}). This permits control of the active heat load delivered to the ion-trap, which was increased in steps of 10~W to a maximum of 80~W. The temperature of the heat sink outlet was monitored until steady-state was reached for a given heater power. This procedure was reproduced separately for Exp 1 and Exp 2 as well as when a heat load was applied to both experiments with the system operated under 23 bar. Figure \ref{fig: Cooling Power} shows an example of the heater power at Exp 2 being incrementally adjusted from 40~W to 80~W while heater power to Exp 1 was held constant at 40~W. With 80~W of applied power, the temperature at Exp 2 remains significantly under 100~K.

The cold head capacity curves provided by the manufacturer \cite{AL330, CH110}, establish the dependency of the cooling power with temperature. Therefore, the total heat load can be calculated from the temperature of the cold heads. By subtracting the known active heat loads applied to the experiment heat sink, the passive heat loads were calculated. The total passive heat load calculated on the cooling system was 86(3)~W when connected to the first experiment and 78(4)~W when only connected to the second. When the two experiments were connected in parallel the total passive heat load was 108(3)~W.

By comparing the results for experiments connected first individually and then in parallel, the contributions to the passive heat load from each section of the cooling circuit can be distinguished. The cryostat vessel is subject to a passive heat load of 56~W, while the first and second experiments (including their circuit loops) contribute 30~W and 22~W, respectively.

\subsection{Helium gas flow rate}\label{sec: Helium mass flow rate}

Understanding the fluid flow dynamics of the single- and two-ion trap experiment is necessary to predict the system performance as it is extended to provide cooling to additional experiments. The total volume flow rate $\dot{V}$ is inversely proportional to the temperature difference $\Delta T$ (K) between the cryostat inlet (T5) and cryofan intake (T3), as described by \cite{Incropera2006}

\begin{equation}
    \dot{V} = \frac{\dot{Q} }{c_{p} \rho_{He} \Delta T},
    \label{eq: He flow rate}
\end{equation}

\noindent where $\dot{Q}$ is the total cooling power (W), $c_p$ is the specific heat capacity of helium at constant pressure (5.517~J/(g$\cdot$K)), and $\rho_{He}$ is the helium density (kg/m$^{3}$). When the cooling system is connected to two experiments in parallel, the maximum volume flow rate, $\dot{V}$, is 0.40~m$^{3}$/hr for a cryofan rotation speed of 21,000~rpm. From the cryofan specifications, the total pressure drop was estimated across the circuit to be 9.8~kPa \cite{Noordenwind}.

However, calculation of the local flow rate at each experiment using equation \ref{eq: He flow rate} was not possible due to measurement offsets. These offsets were attributed to poor thermal anchoring of the Pt100 sensors, which was due to thermal contraction mismatches during cool down. Instead, a calculation of the volume flow rate at a single experiment $\dot{V}_{exp}$ is made via a rearrangement of equation \ref{eq: He flow rate} using two consecutive measurements such that

\begin{equation}
    \dot{V}_{exp} = \frac{\Delta\dot{Q}}{c_p \rho_{exp} (\Delta T_{out} - \Delta T_{in} )},
    \label{eq: He flow at experiment}
\end{equation}

\noindent where $\Delta\dot{Q}$ (W) is the power difference between two consecutive applied heat loads at the ion-trap, $\rho_{exp}$ is the helium density at the experiment location, and $\Delta T_{in}$ (K) and $\Delta T_{out}$ (K), are the steady-state temperature differences measured at the inlet and outlet locations respectively. Using equation \ref{eq: He flow at experiment}, a helium volume flow rate of 0.24~m$^{3}$/hr and 0.16~m$^{3}$/hr is calculated for the first and second experiment respectively. The larger flow rate at Exp 1 is essential since it contains a larger number of heat dissipating components than Exp 2 and therefore requires more cooling power. When combined, these volume flow rates are consistent with the total volume flow rate $\dot{V}$ calculated at the cryostat.

No significant change in the total flow rate was observed during the process of connecting an experiment into the cooling circuit.

\subsection{Vibrations}\label{sec: Vibrations}

\begin{figure}
    \centering
    \includegraphics{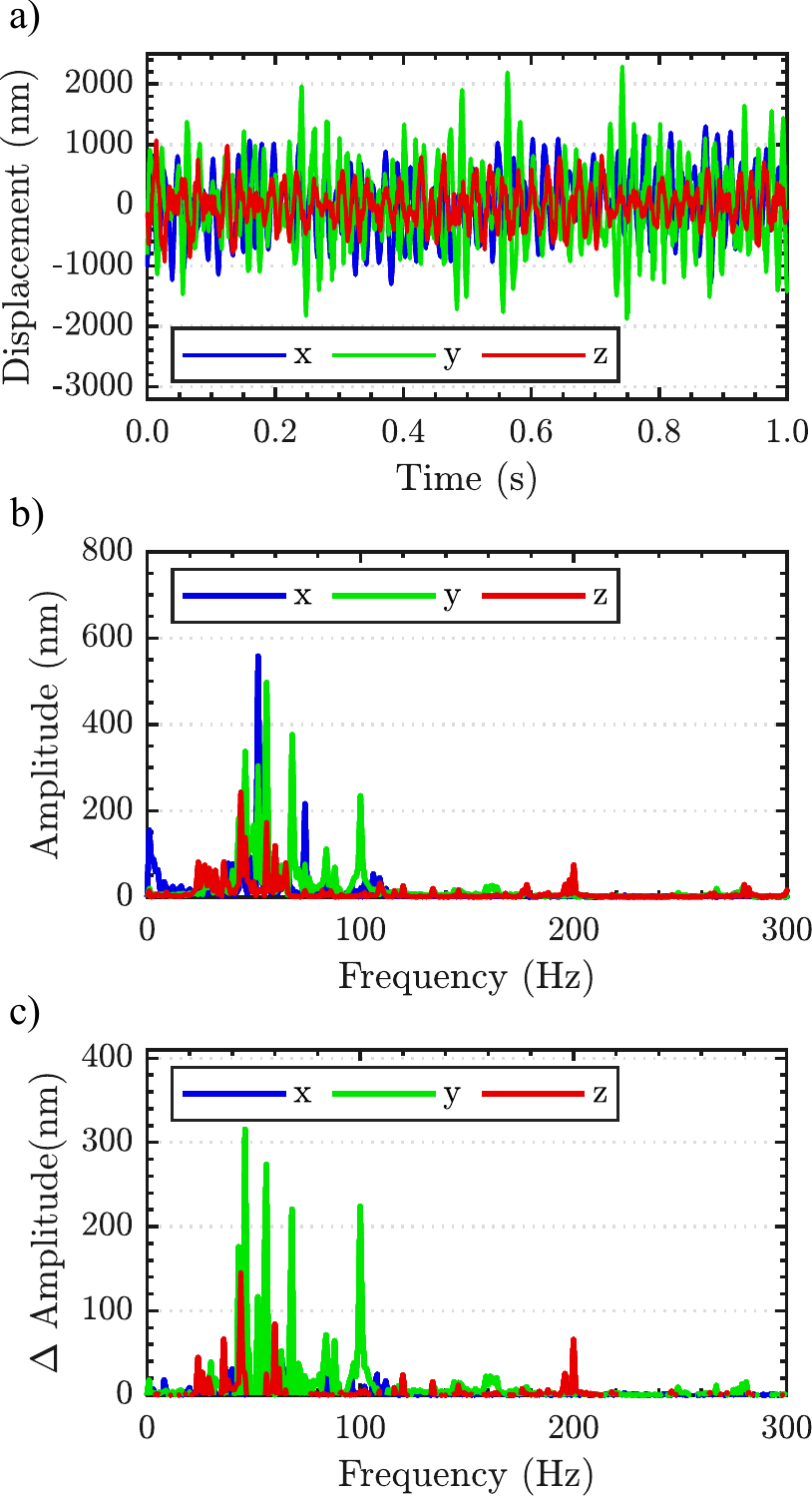}
    \caption{Time evolution (a) and frequency spectrum (b) of the vibration of the in-vacuum heat sink in Exp 1. The x, and z-axis correspond to those shown in Figure \ref{fig: In-Vacuum Mounting}a. In these measurements both cryocooler 1 and 2 are on, the cryogenic fan is on and the heat sink has reached a 40~K steady-state temperature. Measurements of each axis were taken consecutively. c) Amplitude spectrum with background vibrations subtracted.}
    \label{fig: Vibrations}
\end{figure}

While the combination of two GM cryocoolers delivers substantial cooling power at 70~K, the operation of the cryocooler cold heads generates significant vibration, which requires dampening. Measurements with piezo-electric accelerometers (DYTRAN 3143M1) of the cryostat vessel, which houses the cold heads, show maximum displacement occurs at frequencies of 2~Hz and 50~Hz, with amplitudes of 50~$\mu$m and 3~$\mu$m respectively.

To reduce the amplitude of vibrations transmitted from the cryostat vessel to the ion-trap experiments, a number of dampening measures were installed onto the helium transfer lines. The vacuum jacket surrounding the transfer lines is mechanically decoupled from the cryostat via edge-welded bellows. The lines have flexible sections installed at locations close to the helium feedthroughs on each vacuum chamber. Additionally, each vacuum chamber is bolted to an optical table.

Vibration measurements were made at the ion-trap experiment closest to the cryostat, using a heterodyne Michelson interferometer set-up (in a similar fashion to that presented in Ref. \cite{Brandl2016}). Three mirrors were attached to the ion-trap support structure inside the vacuum chamber to reflect 650 nm laser light delivered along three perpendicular axes. Reference mirrors were anchored directly onto the outside of the vacuum chamber. This method permits measurements of vibrations with amplitudes of a few nm and in the frequency range 0--50~kHz.

When the heat sink is operating at 40~K, the measured vibration can be seen in Figure \ref{fig: Vibrations}a which shows the largest vibrations are in the y-direction with a maximum displacement of 2.3~$\mu$m relative to the vacuum chamber. Figure \ref{fig: Vibrations}b shows the frequency spectrum of the vibration as calculated from Figure \ref{fig: Vibrations}a. However, the measurements in Figure \ref{fig: Vibrations}b are the total vibrations experienced by the experiment. To isolate the effect of the cooling system, measurements at 295~K were also taken. Figure \ref{fig: Vibrations}c shows the difference in the amplitudes of vibration between when the cooling system was turned on (Figure \ref{fig: Vibrations}b) and turned off. The most prominent contributions from the cooling system are in the y-direction at 46~Hz, 56~Hz, 68~Hz and 100~Hz with amplitudes of 315~nm, 274~nm, 220~nm and 224~nm respectively.

\section{Scaling from two to four ion-trap experiments}

\begin{table}[h!]
\begin{tabular}{|l|c|}
\hline
\multicolumn{1}{|c|}{{\ul \textbf{Position}}} & {\ul \textbf{Temperature (K)}} \\ \hline
Cryocooler 1 (T1)                             & 51.0                           \\ \hline
Cryocooler 2 (T2)                             & 55.1                           \\ \hline
Fan intake  (T3)                              & 53.0                           \\ \hline
Cryostat outlet  (T4)                         & 55.0                           \\ \hline
Cryostat inlet (T5)                           & 79.5                           \\ \hline
Merge inlet (T9)                              & 58.5                           \\ \hline
Merge outlet (T10)                            & 82.5                           \\ \hline
Exp 1 Inlet (T11)                             & 62.0                           \\ \hline
Exp 1 outlet (T12)                            & 79.0                           \\ \hline
Exp 2 Inlet (T7)                              & 65.0                           \\ \hline
Exp 2 outlet (T8)                             & 70.5                           \\ \hline
\end{tabular}
\caption{Summary of temperature projections for experiment 1 and 2 when configured for four ion-trap experiments. An explanation of the positions labelled in the table can be found in Figure \ref{fig: Cryo System}.}
\label{tab: Table}
\end{table}

The results presented in section \ref{sec: Helium mass flow rate} were used to model the cooling system performance when operating with four experiments connected in parallel. This model assumes a total volume flow rate, $\dot{V}$, of 0.40~m$^{3}$/hr. The passive heat loads are assumed to be 60~W at the cryostat and 30~W for each of the four parallel circuit loops. For the four ion-trap experiments, an active heat load of 75~W is assumed for the first and 12~W each is assumed for the others. These ‘power budgets’ were chosen based on the number of active components within each experiment. The total heat load is therefore 291~W, of which 111~W are active and 180~W are passive. Cryocooler 1 and 2 are expected to operate at 51~K and 55~K, respectively\cite{AL330,CH110}.

At this temperature and under a pressure of 20~bar, the helium gas density at the cryofan is calculated to be 18.16~kg/m$^{3}$. The equivalent mass flow rate at the cryostat is therefore projected to be 2.01~g/s. The valves at each experiment are set to divide the flow rate such that 0.81~g/s of helium is delivered to the 75~W experiment while 0.40~g/s is delivered to each of the three 12~W experiments. Average heat sink temperatures of 65.5~K and 68~K are expected at Exp 1 and Exp 2 respectively. Table \ref{tab: Table} provides a summary of the temperature projections of the model.

\section{\label{sec: Conclusion} Conclusion}

Thermal management is a critical concern within a scalable architecture for ion-trap quantum computing. Presented here is the development of a cryogenic cooling system with high cooling power and a scalable design, applicable to large-scale arrays of ion-traps. This paper demonstrates a laboratory configuration capable of providing 111~W of net cooling power at 70~K to up to four independent ion-trap experiments. By using a remote cryostat that is thermally coupled to a helium gas circulation system, the physical size constraints of existing cryostat systems are no longer an obstacle at the ion-trap location. Therefore, cooling power can be delivered within the constrained footprint area of the ion-trap chips to permit many chips to be closely spaced side by side across a surface. The cooling system therefore satisfies one of the key engineering requirements of the scalable architecture proposed by Lekitsch \textit{et al.}\cite{Lekitsch2017}, and it is a step towards the construction of practical devices for ion-trap quantum information processing. In addition, the system presented here also has applications in other high-power processes that require a scalable distributed cooling network in a similar temperature range. Some examples include high-temperature superconductor technologies \cite{Pamidi2015}, cooling of high-power laser amplifiers \cite{Mason2015} and even in propellant tanks for space exploration \cite{Plachta2018}.

\begin{acknowledgments}
The authors thank Christophe Valahu for providing valuable technical assistance, Bjoern Lekitsch for helpful guidance during planning of the cooling system, Tomas Navickas for his advice on heat sink construction, and Quentin Bodart for the helpful discussions and comments on this manuscript. This work was supported by the U.K. Engineering and Physical Sciences Research Council via the EPSRC Hub in Quantum Computing and Simulation (EP/T001062/1), the U.K. Quantum Technology hub for Networked Quantum Information Technologies (No. EP/M013243/1), the European Commission’s Horizon-2020 Flagship on Quantum Technologies Project No. 820314 (MicroQC), the U.S. Army Research Office under Contract No. W911NF-14-2-0106, the Office of Naval Research under Agreement No. N62909-19-1-2116, the Luxembourg National Research Fund (FNR) (Project Code 11615035), the University of Sussex, and through a studentship in the Quantum Systems Engineering Skills \& Training Hub at Imperial College London funded by the EPSRC (EP/P510257/1).

%This work was supported by:
%\begin{itemize}[noitemsep,topsep=0pt]
%    \item U.K. Engineering and Physical Sciences Research Council via the EPSRC Hub in Quantum Computing and Simulation (EP/T001062/1)
%    \item U.K. Quantum Technology hub for Networked Quantum Information Technologies (No. EP/M013243/1)
%    \item European Commission’s Horizon-2020 Flagship on Quantum Technologies Project No. 820314 (MicroQC)
%    \item  U.S. Army Research Office under Contract No. W911NF-14-2-0106
%    \item  Office of Naval Research under Agreement No. N62909-19-1-2116 and the University of Sussex
%    \item the Luxembourg National Research Fund (FNR) (Project Code 11615035)
%\end{itemize}

\end{acknowledgments}

\section*{Data Availability}

The data that support the findings of this study are available from the corresponding author
upon reasonable request.

%\nocite{*}
\bibliography{Bibliography}% Produces the bibliography via BibTeX.

\end{document}